\documentclass[superscriptaddress,amsmath]{revtex4}
\usepackage{epsfig}

\begin{document}

\title{Chemical Equilibration at the Hagedorn Temperature}

\author{J. Noronha-Hostler}
\email{hostler@th.physik.uni-frankfurt.de} \affiliation{The
International Graduate School for Science (FIGSS), Johann Wolfgang
Goethe--Universit\"at, D-60438 Frankfurt am Main, Germany}
\author{C. Greiner}
\email{carsten.greiner@th.physik.uni-frankfurt.de}
\affiliation{Institut f\"ur Theoretische Physik, Johann Wolfgang
Goethe--Universit\"at, D-60438 Frankfurt am Main, Germany}
\author{I. Shovkovy}
\email{I-Shovkovy@wiu.edu} \affiliation{Western Illinois University, Macomb, IL, 61455 USA}
\begin{abstract}

One important question in relativistic heavy ion collisions is if hadrons, specifically anti-hyperons, are in equilibrium before thermal freezeout because strangeness enhancement has long been pointed to as a signature for Quark Gluon Plasma.  Because anti-baryons have long equilibration times in the hadron gas phase it has been suggested that they are ``born" into equilibrium.   However, Hagedorn states, massive resonances, which are thought to appear near the critical
temperature, contribute to fast chemical equilibration times for a
hadron gas by providing extra degrees of freedom. Here we use master equations to describe the interplay
between Hagedorn resonances, pions, and baryon anti-baryon pairs as
they equilibrate over time and observe if the baryons and anti-baryons  are fully equilibrated within the fireball.
\end{abstract}
\maketitle
\section{Introduction}

It has been suggested that one of the main signatures for the Quark Gluon Plasma (QGP), which is a state of matter that consists of deconfined quarks and gluons, could be the equilibration of anti-hyperons and multi-strange baryons \cite{Koch:1986ud}.  The idea was that within a hadron gas there would not be enough time for these particles to reach chemical equilibrium and thus they should be ``born" into equilibrium after the QGP phase transition \cite{Stock:1999hm,Heinz:2006ur}.  Indeed, experimentally we know that the particle abundancies reach chemical equilibration close to the phase transition \cite{Braun-Munzinger}.  Multi-mesonic reactions within the standard hadron gas model, which can explain abundancies at SPS energies \cite{Rapp:2000gy,Greiner}, cannot account for the short chemical equilibration times and high abundancies of anti-baryons at RHIC \cite{Kapusta,Huovinen:2003sa} (but also see \cite{BSW}).  In this paper we focus on Hagedorn resonances (heavy resonances that have an exponential mass spectrum that appear near the critical temperature) that drive multi-pionic reactions and also produce baryon anti-baryon pairs as was suggested in Ref.\ \cite{Greiner:2004vm}. Once we include the Hagedorn resonances we see that it is possible for the anti-baryons (future work will discuss anti-hyperons and multi-strange baryons) to quickly equilibrate within the fireball just ``below" the phase transition.

Originally, (anti-)strangeness enhancement at CERN-SPS energies in comparison to $pp$-data, which was primarily observed in anti-hyperons and multi-strange baryons, was thought of as a signature for QGP.  Using binary strangeness production reactions such as 
\begin{equation}\label{eqn:strangeprod}
\pi+\bar{p}\leftrightarrow \bar{K}+\bar{\Lambda}
\end{equation}
or binary strangeness exchange reactions 
\begin{equation}\label{eqn:strangeex}
K+\bar{p}\leftrightarrow \pi+\bar{\Lambda}
\end{equation}
it was concluded that it took far too long for chemical equilibrium to be reached within the hadron gas phase \cite{Koch:1986ud}.  Thus, it was proposed that the QGP was already observed at SPS because strange quarks can be produced more abundantly by gluon fusion, which would then account for strangeness enhancement in hadrons following hadronization and rescattering of strange quarks \cite{Koch:1986ud}.

On the other hand, to explain secondary production of anti-hyperons the idea was suggested that strangeness enhancement could be explained using multi-mesonic reactions such as
\begin{equation}\label{eqn:antiproton}
\bar{p}+N\leftrightarrow n\pi
\end{equation}
for the anti-protons \cite{Rapp:2000gy} and for the anti-hyperons
\begin{eqnarray}\label{eqn:antihypall} 
\bar{\Sigma},\bar{\Lambda}+N&\leftrightarrow &n\pi+K\nonumber\\
\bar{\Xi}+N&\leftrightarrow &n\pi+2K\nonumber\\
\bar{\Omega}+N&\leftrightarrow &n\pi+3K
\end{eqnarray}
found in Ref.\ \cite{Greiner}.
The anti-hyperons can be rewritten into the general equation
\begin{equation}\label{eqn:antihyp}
\bar{Y}+N\leftrightarrow  n\pi+n_{Y}K.
\end{equation}
The arrows in Eqs.\ (\ref{eqn:antiproton}), (\ref{eqn:antihypall})
signify the equal probability that a decay can occur in either
direction otherwise known as detailed balance.

The time scale of a standard hadron gas at SPS can be then estimated using the multi mesonic reactions in Eq.\ (\ref{eqn:antiproton}), (\ref{eqn:antihypall}).  Using
\begin{equation}\label{eqn:crosssection}
    \sigma_{N\bar{Y}}\approx\sigma_{N\bar{p}}\approx 50\;\mathrm{mb}
\end{equation}
the chemical equilibration time is then
\begin{equation}\label{eqn:prochemtime}
    \tau_{\bar{Y}}=\left(\Gamma_{\bar{Y}}\right)^{-1}=
    \frac{1}{\langle\langle\sigma_{N\bar{Y}}v_{N\bar{Y}}\rangle\rangle
    \rho_{B}}\approx 1-3\; \frac{\mathrm{fm}}{c},
\end{equation}
where $\rho_{B}\approx \rho_{0}\;\mathrm{to}\;2\rho_{0}$, which is typical for SPS \cite{Greiner,Rapp:2000gy}.
Therefore, the time scale given in Eq.\ (\ref{eqn:prochemtime}) is
short enough to account for chemical equilibration within the cooling hadronic
fireball at SPS.

\begin{figure}[h]
\begin{center}
\leavevmode
\includegraphics[width=6cm]{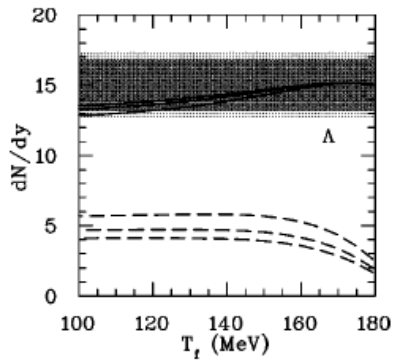}
\end{center}
\caption{Anti-lambda production in the most central collisions at
RHIC.   The bottom three dashed lines start with no anti-lambdas at
$T=180$ MeV whereas the solid lines assume that the anti-lambdas
begin in equilibrium.  The shaded area shows the experimental
results taken from Ref.\ \cite{Huovinen:2003sa}.  The set of lines shows the difference for varying coupling coefficients.} \label{fig:alambda}
\end{figure}

If we apply our same understanding of the hadron gas phase to RHIC temperatures our time scales are much longer.  The equilibrium rate of $\Omega$ at RHIC at $T=170$ MeV is
$\Gamma_{\Omega}^{chem}\approx\langle\sigma_{\Omega
\bar{B}}v_{\Omega \bar{B}}\rangle N_{\bar{B}}$ where the cross section is $\sigma\approx
30\;\mathrm{mb}$ and the baryon density is
$N_{B}^{eq}=N_{\bar{B}}^{eq}\approx0.04\;\mathrm{fm}^{-3}$ , which leads to a time scale of
$\tau_{\Omega}=\frac{1}{\Gamma_{\Omega}}\approx
10\;\frac{\mathrm{fm}}{\mathrm{c}}$. However, considering
that the fireball's time scale in the hadronic stage is $\tau<
4\;\frac{\mathrm{fm}}{\mathrm{c}}$, a standard hadron
gas could not explain the apparent chemical equilibration observed in
baryons within the fireball.  Moreover, these results were also backed up in Ref.\ \cite{Kapusta} where using a fluctuation-dissipation theorem it was found that the equilibration time of baryons and anti-baryons, $\tau\approx
10\left[\frac{\mathrm{fm}}{\mathrm{c}}\right]$, at RHIC temperatures. 
In the $5\%$ most central Au-Au collisions the baryon anti-baryon production is roughly
three  times lower than the measured experimental values if it 
starts out of equilibrium (specifically at zero for reactions of type (\ref{eqn:antiproton}) and (\ref{eqn:antihypall})) shown in  Fig.\ (\ref{fig:alambda}), taken from
Ref.\ \cite{Huovinen:2003sa}, for $\bar{\Lambda}$ production.
Because of the apparent differences in the equilibration  times some
have suggested that the hadrons are ``born" into equilibrium i.e.
the system is already in a chemically frozen out state at the end of
the phase transition \cite{Heinz:2006ur}.

In this paper we
use Hagedorn states to provide the extra degrees of
freedom needed to match experimental results when the anti-baryons
begin out of equilibrium.  Baryon anti-baryon production develops according to the possible reaction
\begin{equation}\label{eqn:bbproduction}
n\pi\leftrightarrow HS\leftrightarrow n\pi+B\bar{B},
\end{equation}  
which is initially discussed in Ref.\ \cite{Greiner:2004vm}.  The Hagedorn states arethought as  massive resonances with short time scales and only contribute near the critical temperature.  They can then catalyze rapid equilibration of baryons and anti-baryons near $T_{c} $.  The results obtained here lead us to believe that the baryon anti-baryon pairs have sufficient time to equilibrate within the fireball close to the phase boundary. 

\section{Hagedorn States}

In the 1960's Hagedorn found a fit for an experimentally growing
mass spectrum \cite{Hagedorn:1968jf}, $\rho$, which is described as
\begin{eqnarray}\label{eqn:Hagedorn}
    \rho&=&\int_{m_{0}}^{M}F(m)e^{\frac{m}{T_{H}}}dm,\nonumber\\
    F(m)&=&\frac{A}{\left(m^2 +m^2 _{0}\right)^{\frac{5}{4}}}.
\end{eqnarray}
where the minimum mass is $m_{0}=500$ MeV, the
maximum mass is $M=7$ GeV, the Hagedorn temperature is $T_{H}=180$
MeV, which fits within Lattice QCD predictions \cite{Karsch}, and $A=0.5$ is a free parameter.  Because of the exponential
growth, near the Hagedorn temperature the Hagedorn states can
account for the extra degrees of freedom needed to match
experimental values.  $A$ is then chosen by looking at the energy
density and trying take into account the extra degrees of freedom
needed. Here we are considering only mesonic, non-strange Hagedorn
states with masses between $2\;\mathrm{GeV\; and}\;7$ GeV.

To describe the dynamical behaviour of the Hagedorn states we use rate
equations.  Rate equations have both loss and gain terms and have the basic form
\begin{equation}\label{eqn:rateex}
    \frac{dn}{dt}=-loss+gain.
\end{equation}
We assume a system with no net baryon density i.e. $N_{B}=N_{\bar{B}}=N_{B\bar{B}}$.  This should approximately be the case at RHIC at midrapidity.  For the reaction $HS\leftrightarrow n\pi+B\bar{B}$ the behaviour of
the density of the $i^{th}$ Hagedorn resonance $N_{R(i)}$, pions
$N_{\pi }$, and baryon anti-baryon pairs $N_{B\bar{B}}$ is described
 by the following set of equation:
\begin{eqnarray}\label{eqn:networkbab}
\frac{dN_{R(i)}}{dt}&=&-\Gamma_{i}^{tot}
N_{R(i)}+\sum_{n}\Gamma^{tot}_{i,\pi} \Re _{i,n}(T) (N_{\pi
})^{n}B_{i\rightarrow n\pi}\nonumber\\
&+&\Gamma_{i,B\bar{B}}^{tot} \Re_{i,\langle n\rangle}^{\langle n\rangle\pi B\bar{B}} (T)
(N_{\pi })^{\langle n\rangle} N_{B\bar{B}}^2 {}\nonumber\\
\frac{dN_{\pi }}{dt} &=&\sum_{i} \sum_{n}\Gamma_{i,\pi}^{tot} n
B_{i\rightarrow n\pi}\left(N_{R(i)}-\Re (T) (N_{\pi })^{n}
\right)\nonumber\\
&+&\sum_{i} \Gamma_{i,B\bar{B}}^{tot} \langle n\rangle\left(N_{R(i)}-
\Re_{i,\langle n\rangle}^{{\langle n\rangle}\pi B\bar{B}} (T)
(N_{\pi })^{\langle n\rangle}N_{B\bar{B}}^2\right){} \nonumber\\
\frac{dN_{B\bar{B}}}{dt}&=&-\sum_{i}\Gamma_{i,B\bar{B}}^{tot}\left(
N_{B\bar{B}}^2 N_{\pi}^{\langle n\rangle} \Re _{i,\langle n\rangle}(T)- N_{R(i)}\right)
\end{eqnarray}
where $\Gamma$ is the decay width, $B_{i\rightarrow n\pi}$ 
represents the branching ratios, and $\langle n\rangle$ is the average number of pions that each Hagedorn state decays into when a baryon anti-baryon pair is included. We also have two separate detailed balance factors, $\Re (T)=\frac{N_{R(i)}^{eq}}{\left(N_{\pi}^{eq}\right)^{n}}$ for the decay $HS\leftrightarrow n\pi$ and  $\Re_{i,\langle n\rangle}^{{\langle n\rangle}\pi B\bar{B}} (T)=\frac{N_{R(i)}^{eq}}{N_{B\bar{B}}^2 \left(N_{\pi}^{eq}\right)^{n}}$  for the decay $HS\leftrightarrow n\pi+B\bar{B}$.  The detailed balance factors ensure that detailed balance is maintained in equilibrium.  They are also temperature dependent because the equilibrium values of the density are dependent on the temperature.

The branching ratios are described by a gaussian distribution
\begin{eqnarray} \label{eqn:branchingratio}
B_{i\rightarrow n\pi}\approx
\frac{1}{\sigma\sqrt{2\pi}}e^{-\frac{(n-\langle n\rangle)^{2}}{2\sigma ^{2}}}
\end{eqnarray}
where $\langle n\rangle=0.6+0.3\frac{m_{i}}{m_{\pi}}\approx 5-16$ is the average pion number that each
Hagedorn state decays into and $\sigma=0.26\sqrt{\frac{m_{i}}{m_{\pi}}}$ is the width of the
distribution \cite{Hamer:1972wz}.   The decay width $\Gamma_{i}=0.17m_{i}-88$, which has the range $\Gamma_{i}=250\;\mathrm{MeV\;to}\;1100$ MeV, is a linear fit extrapolated from
the data from the particle data group \cite{Eidelman:2004wy}.  In the future we will also use a microcanonical model to find the branching ratios as shown in Ref. \cite{Liu}.  

\section{Results}

We have solved the rate equations in Eq.\ (\ref{eqn:networkbab}) considering several different initial conditions for a statistical system.  At first we take the simplest example and observe only the decay $HS\leftrightarrow n\pi$ when the pions are held in equilibrium.  We then consider a resonance bath and allow the pions to equilibrate.  We also allow both the pions and the resonances to develop until they reach equilibrium. Then we include baryon anti-baryon pairs into our decay $HS\rightarrow  n\pi+B\bar{B}$ where at first the pions are  held in equilibrium, then the Hagedorn states, and also both the Hagedorn resonances and the pions are held in equilibrium while the baryon anti-baryon pairs are allowed to equilibrate.  Finally, we consider the case when all the constituents are allowed to equilibrate simultaneously.  

\subsection{Case 1 $HS\leftrightarrow n\pi$: Pions held in Equilibrium}

Initially, when the Hagedorn states decay only
into multiple pions, $HS\leftrightarrow n\pi$ we assume that
the pions start in equilibrium and the Hagedorn resonances start at
zero, which could be an approximation for a physical system immediately following hadronization when the correlation lengths are very short. Then we can make an estimate for the time scale with the
inverse of the decay width $\Gamma=\frac{1}{\tau}$.  The
equilibration time estimate is then between
$\tau\approx0.2\;\frac{\mathrm{fm}}{c}\mathrm{\;and}\;0.5\;\frac{\mathrm{fm}}{c}$.
\begin{figure}[h!]
\begin{center}
\includegraphics[width=6cm,angle=270]{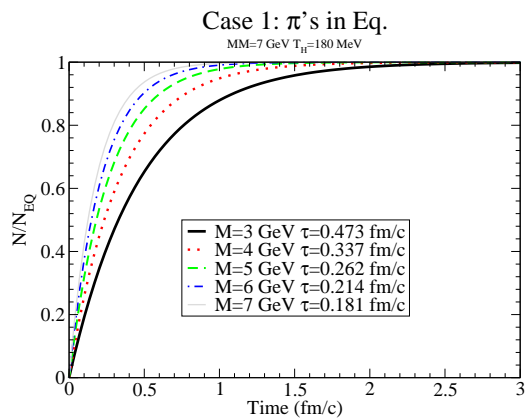}
\end{center}
\caption{Hagedorn resonances for various masses at $T=170$ MeV.}
\label{fig:pionsineq}
\end{figure}
In Fig.\ (\ref{fig:pionsineq}) the result of the rate equations is
shown, which matches the time estimates well.  From Fig.\ (\ref{fig:pionsineq}) we see that the Hagedorn states equilibrate quickly (in comparison to typical expansion times).  Hence, they should be at chemical equilibrium as long as the pions are at chemical equilibrium.

\subsection{Case 2 $HS\leftrightarrow n\pi$: Hagedorn Resonances held in Equilibrium}

\begin{figure}[h!]
\begin{center}
\includegraphics[width=6cm,angle=270]{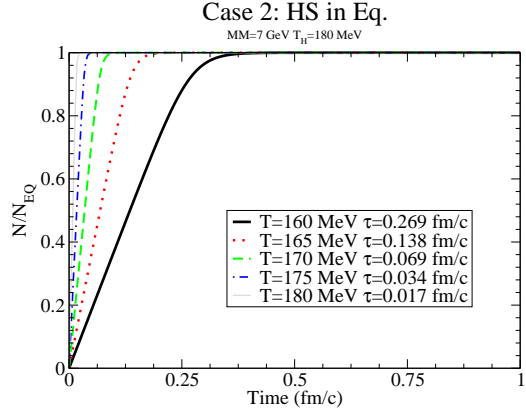}
\end{center}
\caption{Number density of pions as they go to equilibrium while the
Hagedorn resonances are held in equilibrium. } \label{fig:reineq}
\end{figure}

On the other hand, if the Hagedorn resonances are held in
equilibrium and treated like a resonance bath then we can define a
new effective production rate for pions:
\begin{equation}\label{eqn:gamHSinEq}
    \Gamma_{\pi}^{eff}=\sum_{i}\Gamma_{i}^{tot}\langle n_{\pi}^{i}\rangle\frac{N_{R(i)}^{eq}}{N_{\pi}^{eq}},
\end{equation}
which gives the range for the equilibration time
$\tau\approx0.02\;\frac{\mathrm{fm}}{c}\mathrm{\;to}\;0.25\;\frac{\mathrm{fm}}{c}$. Here the pions start at
zero, which could be an approximation for a physical system immediately following hadronization when the correlation lengths are very long.  Again the time scale can be compared with the results from
the rate equations, which is shown in Fig.\ (\ref{fig:reineq}).
Indeed, the pion become populated very quickly towards chemical equilibrium for all temperatures.  

\subsection{Case 3 $HS\leftrightarrow n\pi$: Hagedorn Resonances and Pions are both out of Equilibrium}

The most interesting  case is when both the Hagedorn states
and the pions are out of equilibrium.  The coupled network of rate
equations is
\begin{eqnarray} \label{eqn:setresonances}
\frac{dN_{R(i)}}{dt}&=&\Gamma^{tot}_{i} \left( \sum_{n=2}
\Re (T) (N_{\pi })^{n}B_{i\rightarrow n\pi}-N_{R(i)}\right) {} \nonumber\\
\frac{dN_{\pi }}{dt} &=&\sum _{i}^{\infty} \sum _{n=2}^{\infty}
\Gamma_{i}^{tot} n B_{i\rightarrow n\pi}(N_{R(i)}-
  \Re (T) (N_{\pi })^{n} ).
\end{eqnarray}
The right-hand side of the rate equations goes to zero when the
particles are in equilibrium.  In Eq.\ (\ref{eqn:setresonances}) the
resonances and pions reach a quasi-equilibrium configuration before full
equilibrium is reached.  In the quasi-equilibrium state the right-hand side nears zero and thus slows down the equilibration time, thus, $N_{R(i)}\approx N_{R(i)}^{eq}\left(\frac{N_{\pi}}{N_{\pi}^{eq}}\right)^{\langle n\rangle}$ and $N_{\pi}\approx N_{\pi}^{eq}\left(\frac{N_{R(i)}}{N_{R(i)}^{eq}}\right)^{\frac{1}{\langle n\rangle}}$. 
\begin{figure}[h]
\centering 
\begin{tabular}{ccc}
\epsfig{file=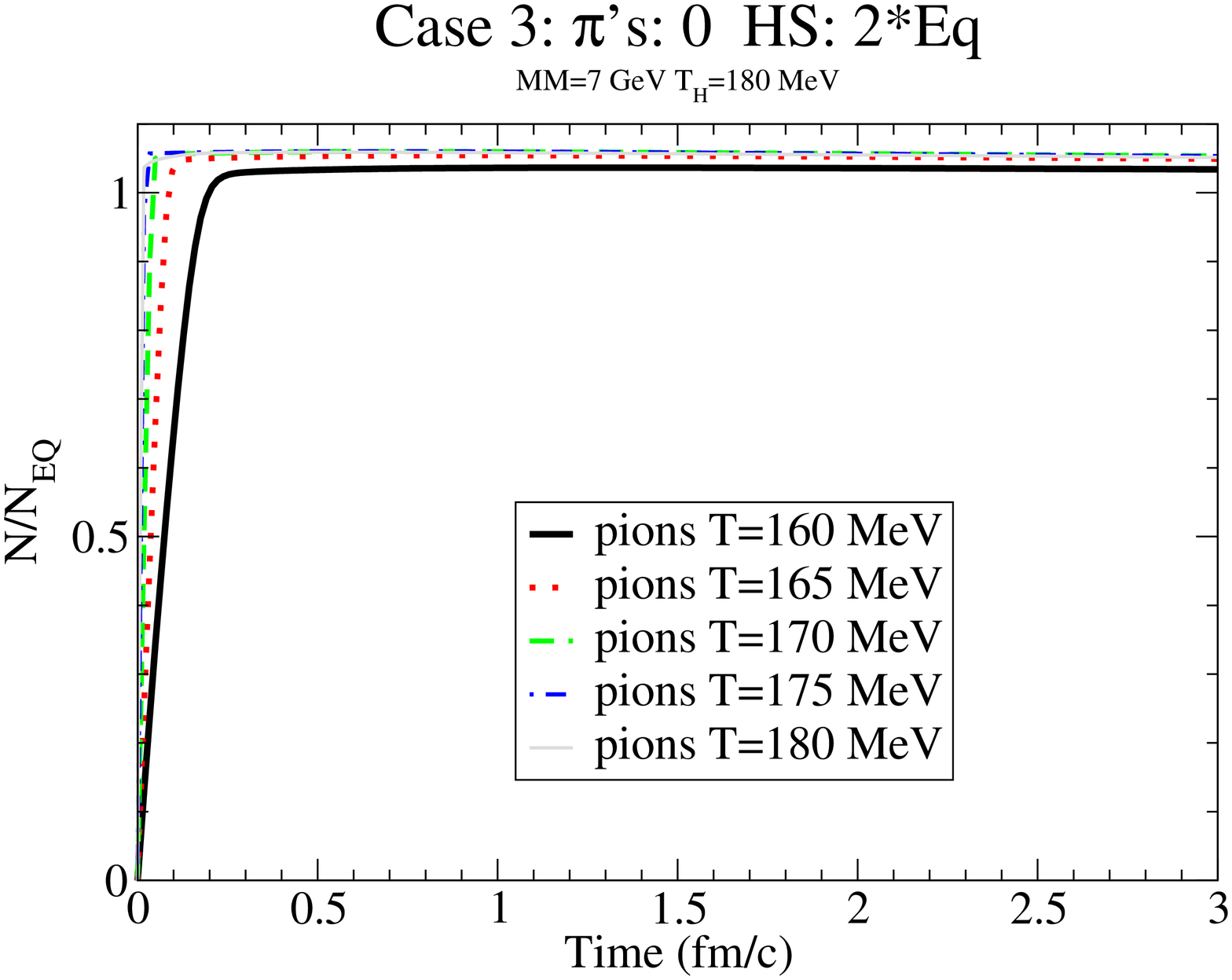,width=0.255\linewidth,angle=270,clip=} &
\epsfig{file=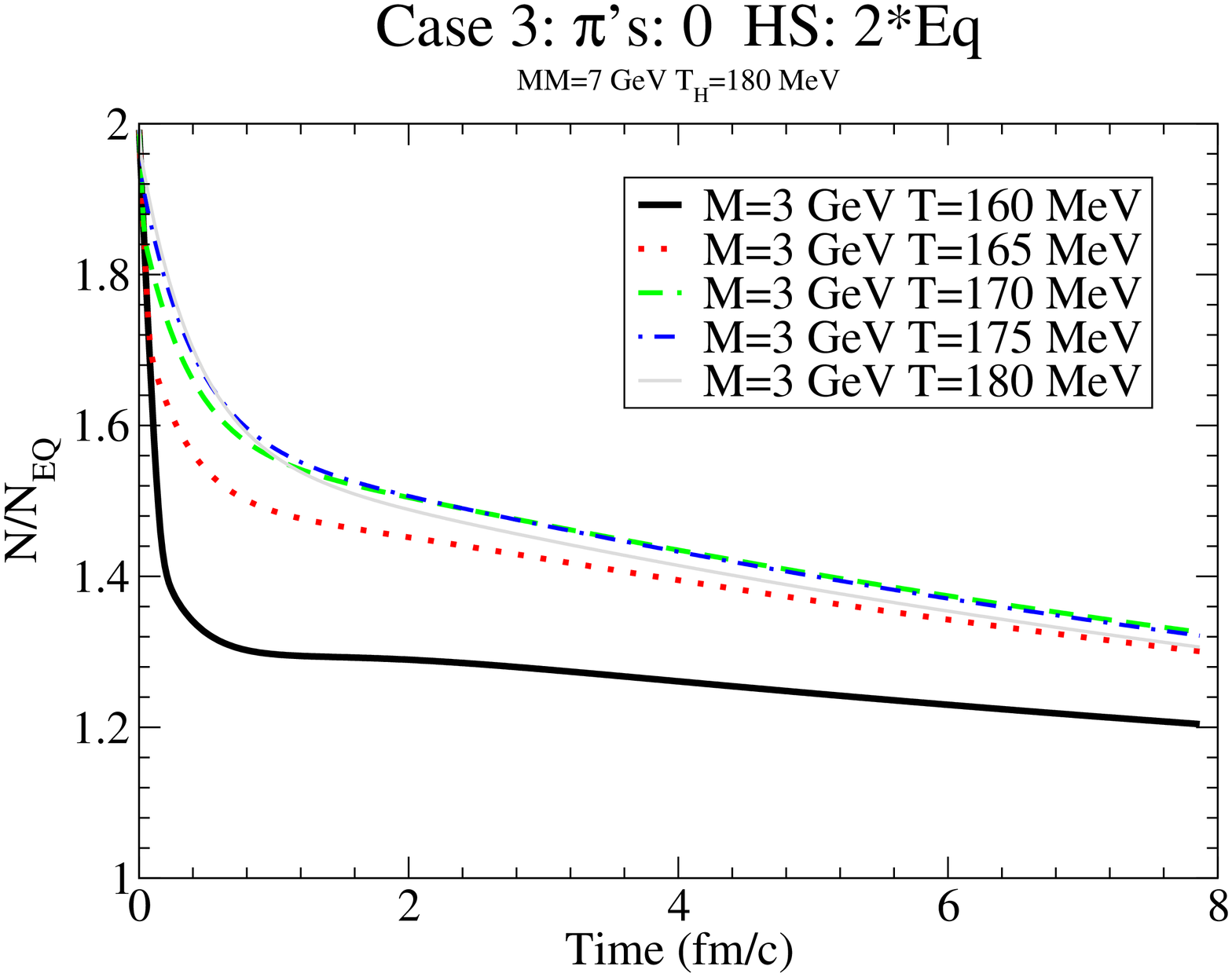,width=0.255\linewidth,angle=270,clip=} &
\epsfig{file=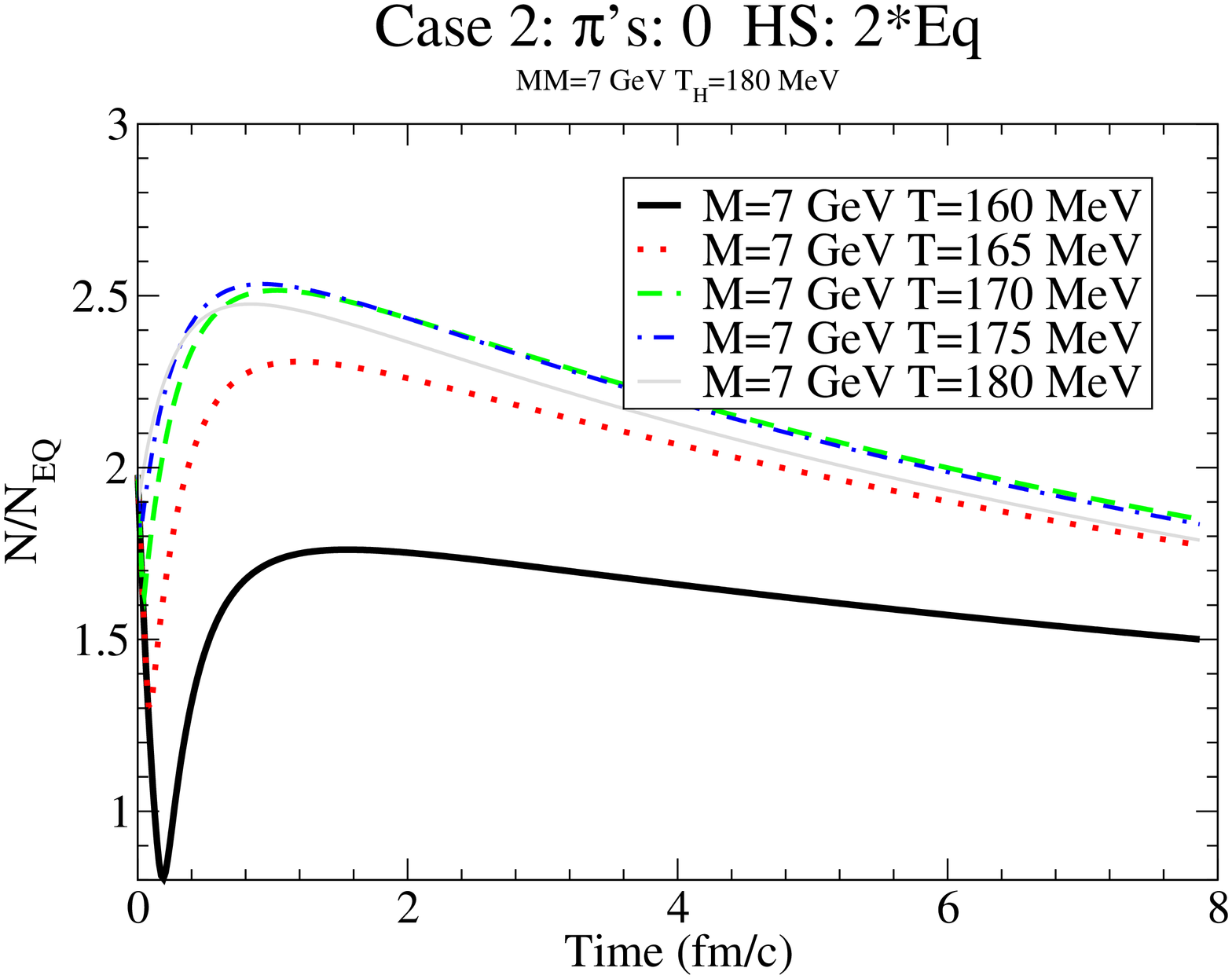,width=0.255\linewidth,angle=270,clip=} \\
\end{tabular}
\caption{Here the resonances are at twice the equilibrium values and
the pions start at zero.} \label{fig:2re0pi}
\end{figure}
In Fig.\ (\ref{fig:2re0pi}) quasi-equilibrium is quickly reached on
the time scales from case 1 and case 2.

The quasi-equilibrium can be understood by looking at the number of effective pions in the system i.e.
\begin{equation}\label{eqn:effpionspre}
 \tilde{N}_{\pi}=N_{\pi}+\sum_{R}\langle n\rangle N_{R}
\end{equation}
where $\langle n\rangle$ is the average number of pions that the Hagedorn resonance $N_{R}$ will decay into.
Essentially in this particular example, there are too many effectives pions in the system and they must be killed off, which can account for the longer time scales.  The direct pions are practically already in equilibrium after $\tau>0.5\;\frac{\mathrm{fm}}{c}$, especially at higher temperatures, however, the Hagedorn states are clearly overpopulated, which implies that the effective number of pions are also overpopulated.  Thus, the long time scale consists of reactions such as such as $n\pi\rightarrow HS\rightarrow m\pi$ where $n>m$, which kills off the effective pions in the system.  Moreover, the heavier Hagedorn resonances are more overpopulated because they have a higher probability to decay into more pions than lighter resonances.  A more in depth look into the time scales of the effective pion number will be discussed in an upcoming paper.

\subsection{$HS\leftrightarrow n\pi+B\bar{B}$}

Finally, we want to determine the production of baryon anti-baryon pairs close to the phase transition.  The baryon anti-baryon pairs are produced through the reaction found in Eq.\ (\ref{eqn:bbproduction}) and these particular branching ratios are determined using a microcanonical model.   The average pion number for
each corresponding mass is calculated by making a fit to the
multiplicities in the  microcanonical model found in Ref.\
\cite{Liu}.
\begin{figure}[h!]
\begin{center}
\includegraphics[width=8cm]{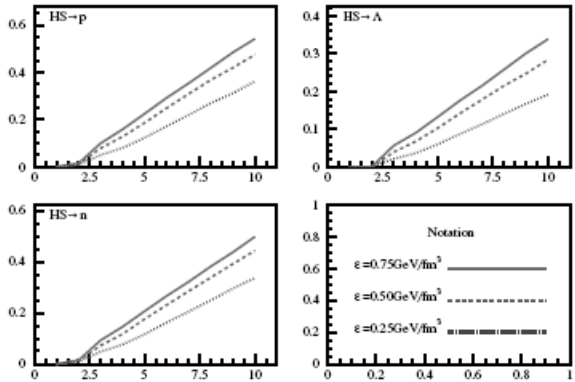}
\end{center}
\caption{Various baryon multiplicities, which are used to determine the total baryon density in our system, for the decay $HS\leftrightarrow n\pi +B\bar{B}$
 \cite{Greiner:2004vm,Liu}.} \label{fig:micro}
\end{figure}

For baryon anti-baryon pairs we need to also reconfigure the decay
widths, which is also done with the baryon multiplicities from Ref.\
\cite{Liu}.  We assume that for every
baryon there is a corresponding anti-baryon and then we add up the
multiplicities for the three dominat baryons: the proton, the neutron
and lambda, to determine the total baryon multiplicity $\langle
B\rangle$ as shown in Fig.\ (\ref{fig:micro})
We can then estimate the relative decay width for the baryon anti-baryon
decay, $\Gamma_{i,B\bar{B}}^{tot}=\langle
B\rangle\Gamma_{i,\pi}^{tot}$.  The total baryon multiplicity varies
between $0.2\mathrm{\;and}\;0.4$, which then gives a decay width between
$\Gamma_{i,B\bar{B}}^{tot}=50\;\mathrm{MeV\;and}\;450$ MeV \cite{Greiner:2004vm}.

\begin{figure}[h!]
\centering
\begin{tabular}{ccc}
\epsfig{file=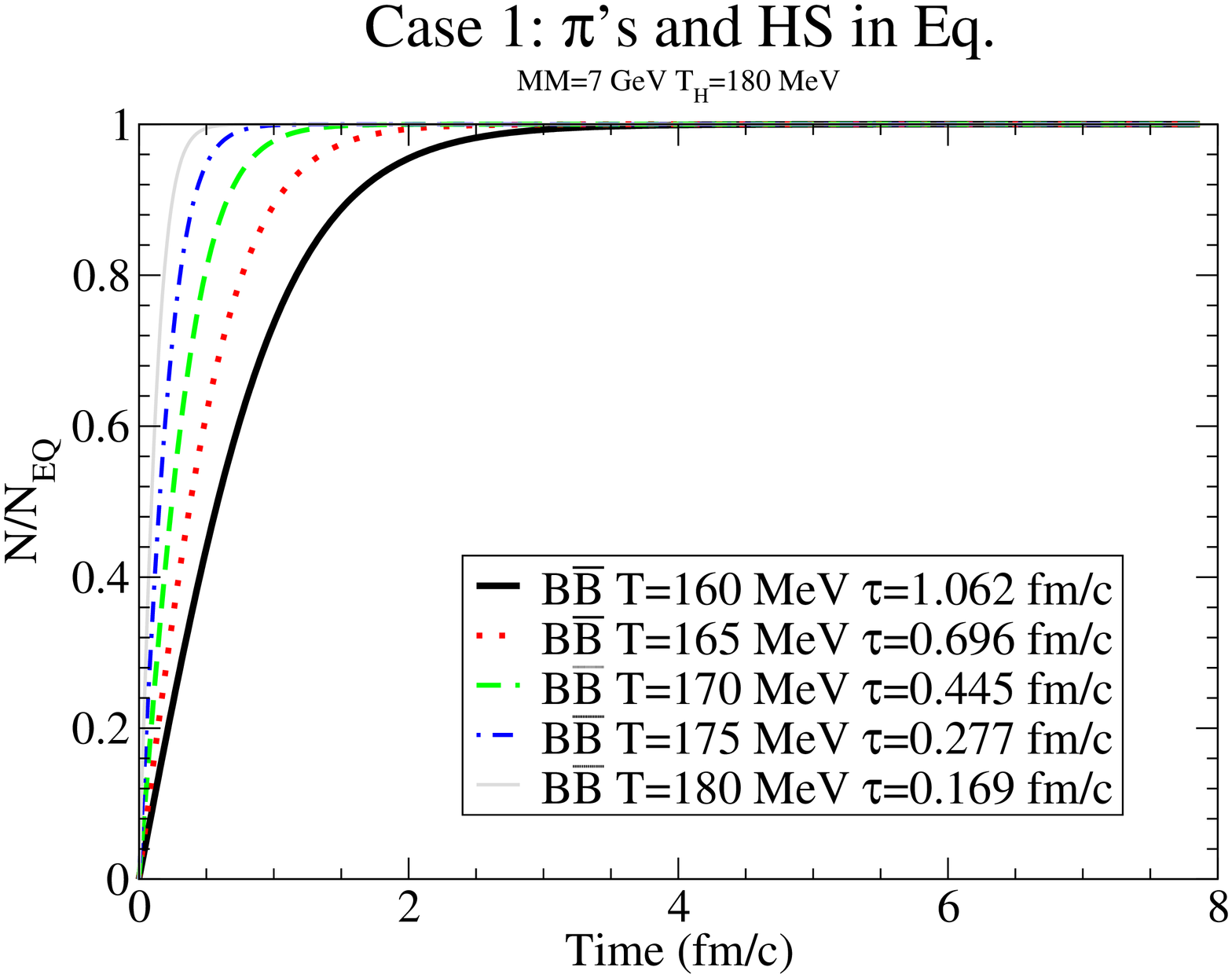,width=0.255\linewidth,angle=270,clip=} &
\epsfig{file=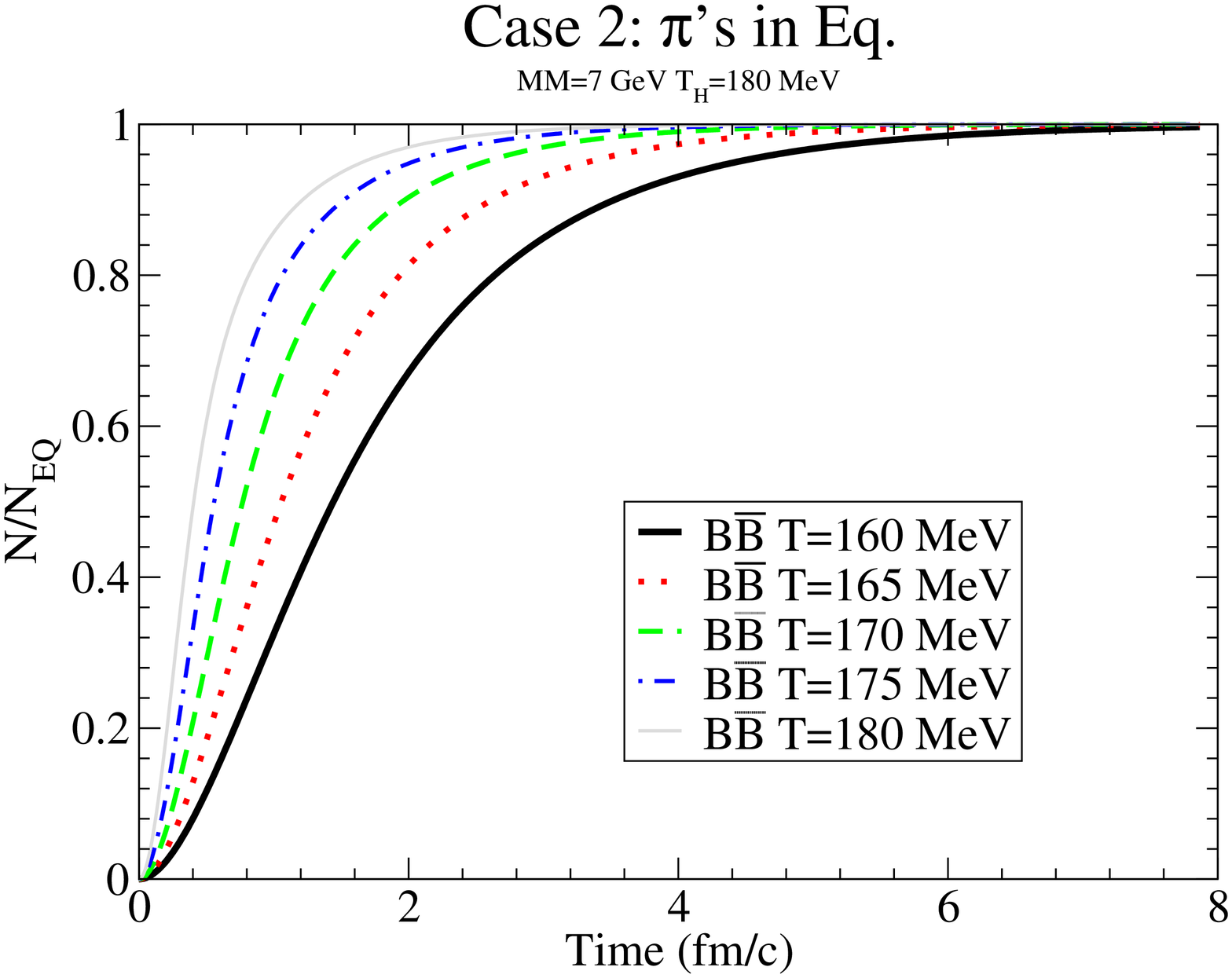,width=0.255\linewidth,angle=270,clip=} &
\epsfig{file=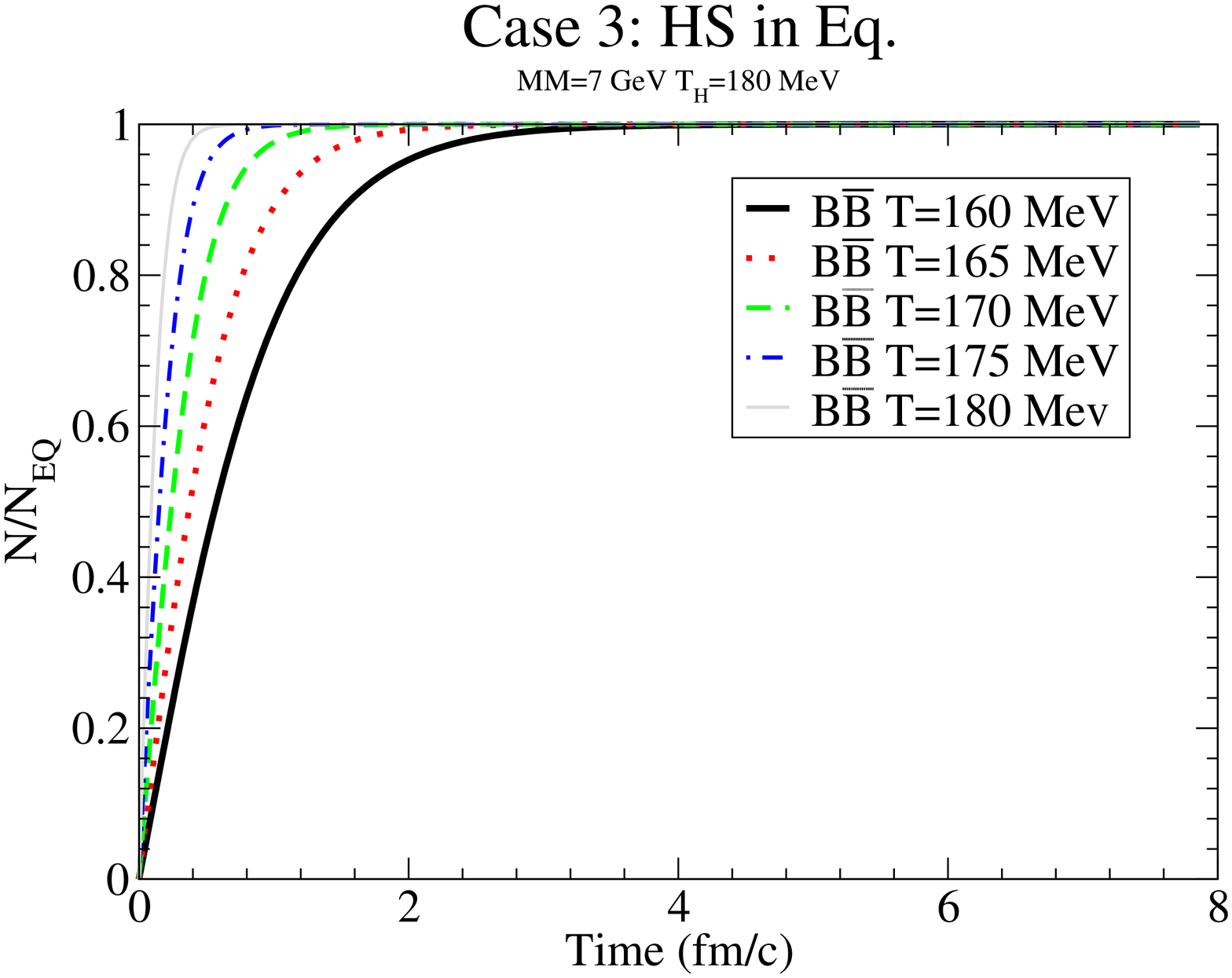,width=0.255\linewidth,angle=270,clip=} \\
\end{tabular}
\caption{Case 1: Solution of the baryon anti-baryon rate equation when the
pions and Hagedorn states are held in equilibrium.  Case 2: Baryon anti-baryon
pairs when the pions are held in equilibrium.  Case 3: Baryon anti-baryon pairs when
the Hagedorn states are held in equilibrium.}
\label{fig:INeqBaB}
\end{figure}

We start with case 1 when both the Hagedorn resonances and the pions can be held in
equilibrium, while the baryon anti-baryon pairs are driven to
equilibrium as shown in case 1 in Fig.\ (\ref{fig:INeqBaB}).
Case 2 is again when the pions are held in
equilibrium (and the resonances begin at zero) and, case 3 is when the Hagedorn resonances are held in equilibrium (and the pions are started at zero).  As
before we can make a time scale estimate, this time for the baryon
anti-baryon pairs. Our effective chemical equilibration  for the baryon anti-baryon pairs is then
\begin{equation}\label{eqn:babdw}
\Gamma_{i,B\bar{B}}^{eff}=-\sum_{i}\Gamma_{i,B\bar{B}}^{tot}
   \left(\frac{ N_{R(i)}^{eq}}{N_{B\bar{B}}^{eq}}\right),
\end{equation}
which gives equilibration times
$\tau=0.2\;\frac{\mathrm{fm}}{c}\mathrm{\;to}\;1\;\frac{\mathrm{fm}}{c}$.  The results of the pions
being held in equilibrium are shown in case 2 in Fig.\ (\ref{fig:INeqBaB})
and the results for the Hagedorn states held in equilibrium are
shown in case 3 in Fig.\ (\ref{fig:INeqBaB}).

For case 2 the baryon anti-baryon pairs take slightly longer to equilibrate than their time scale estimate because quasi-equilibrium is reached with the Hagedorn states, which occurs around $\tau\approx 1\;\frac{fm}{c}$.  Case 3 is not as affected by a quasi-equilibrium state because the pions equilibrate so quickly that they do not affect the equilibration time of the baryon anti-baryon pairs. Comparing the graph of baryon anti-baryon pairs in case 1 to that in case 2 in Fig.\ (\ref{fig:INeqBaB}) we clearly see that they are still almost identical.  The reason is that the baryon anti-baryon pairs are not affected by the pions because the pions equilibrate almost immediately and thus the approximation that the pions are held in equilibrium can be made.

\begin{figure}[h!]
\centering
\begin{tabular}{cc}
\epsfig{file=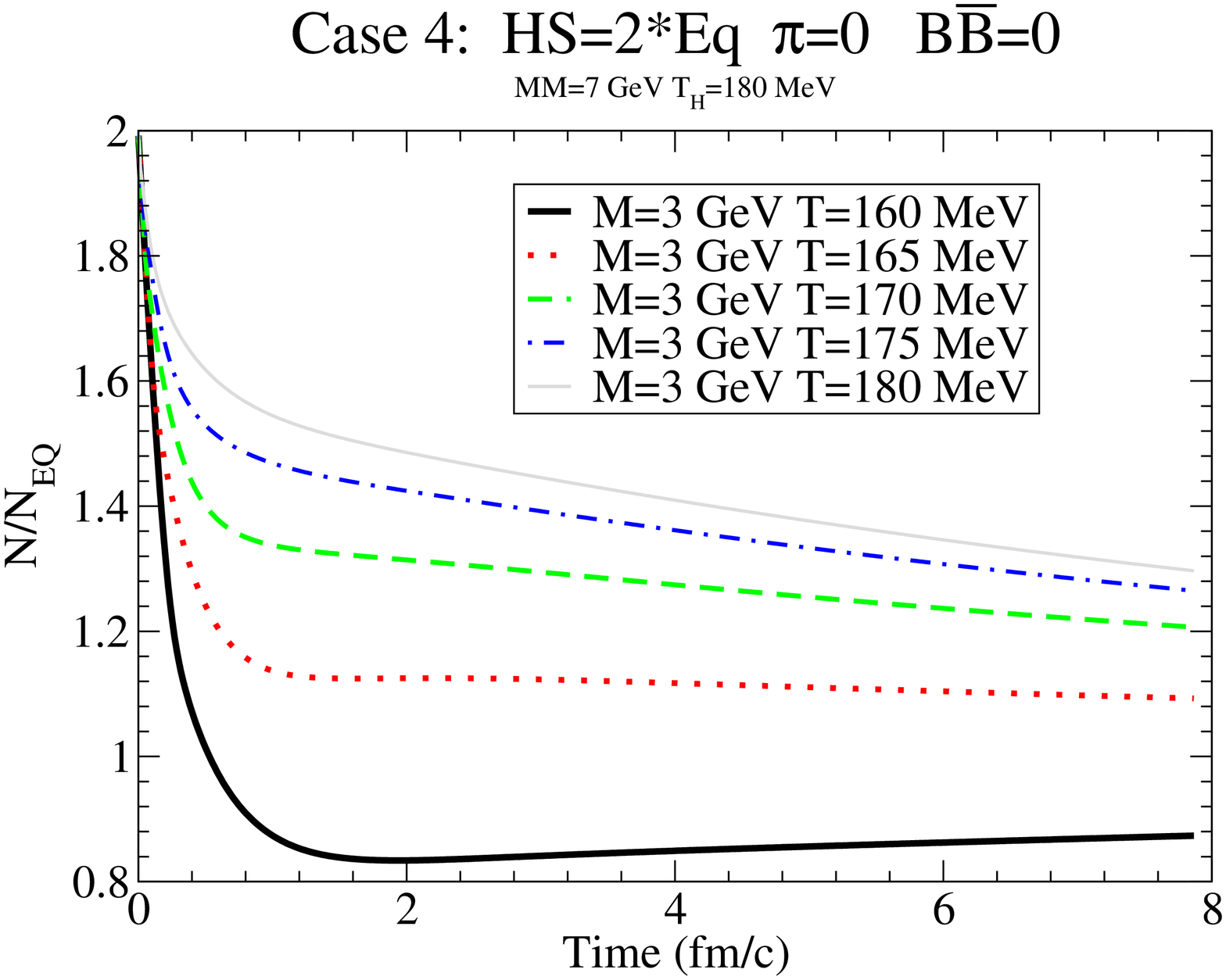,width=0.3\linewidth,angle=270,clip=} &
\epsfig{file=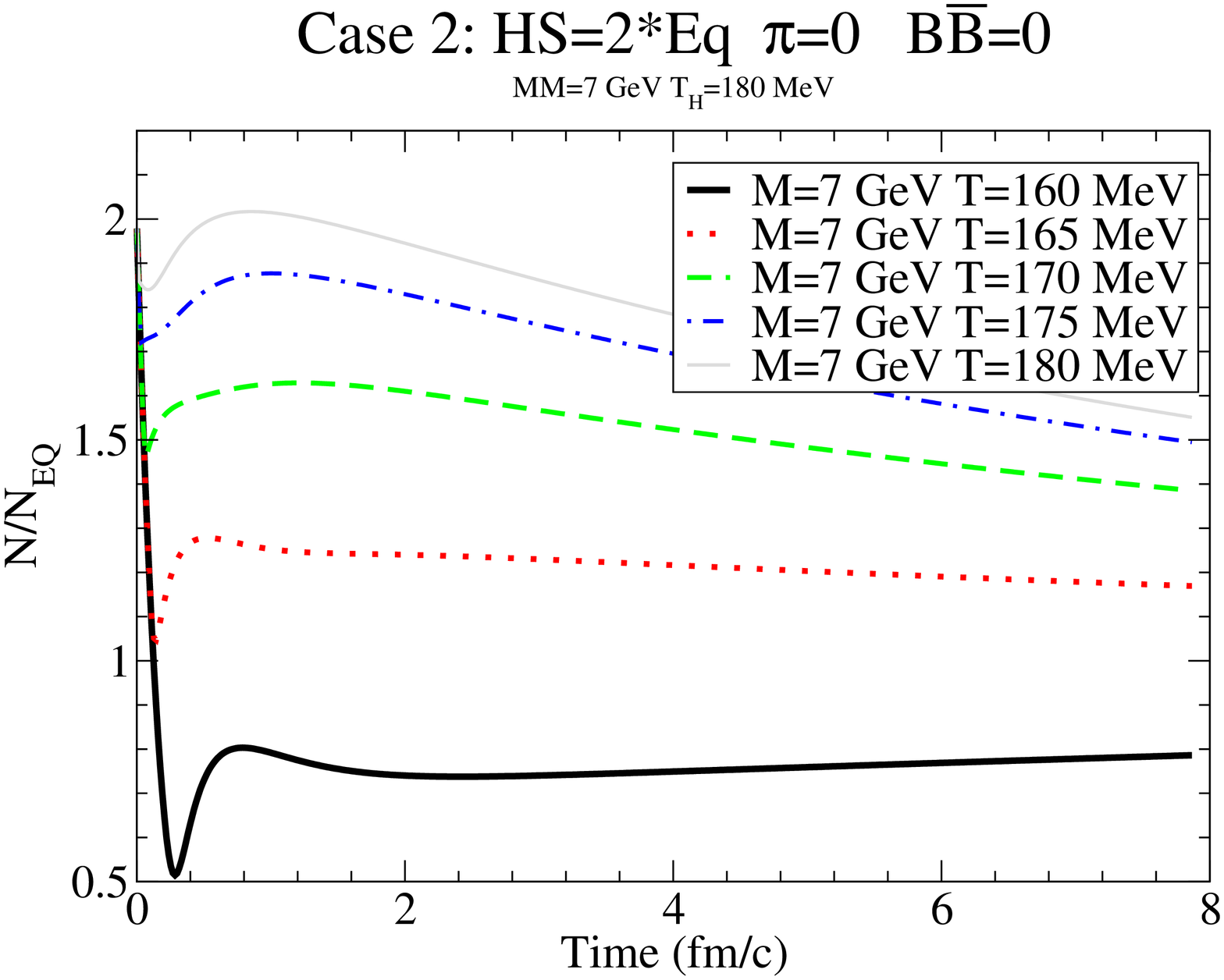,width=0.3\linewidth,angle=270,clip=} \\
\epsfig{file=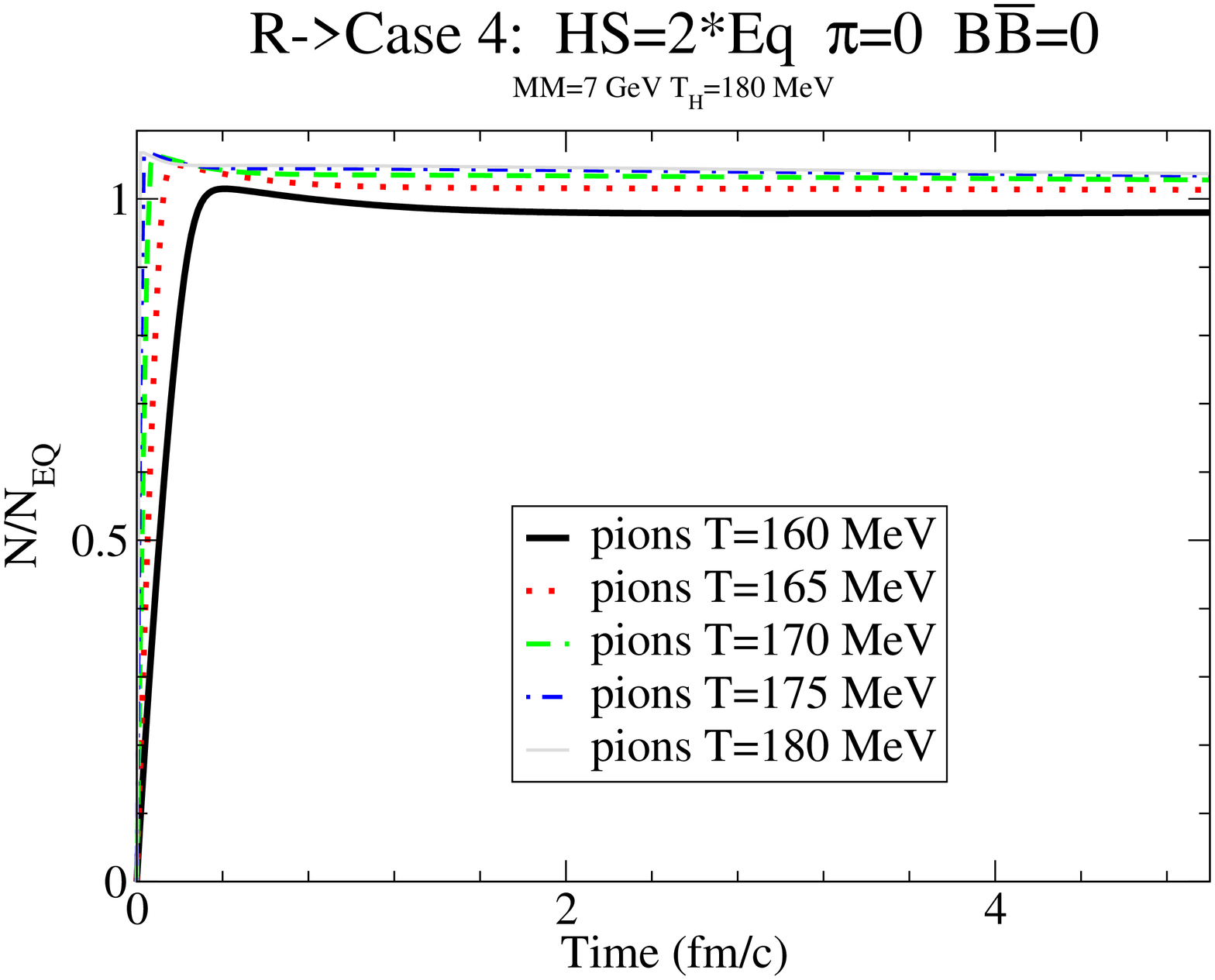,width=0.3\linewidth,angle=270,clip=} &
\epsfig{file=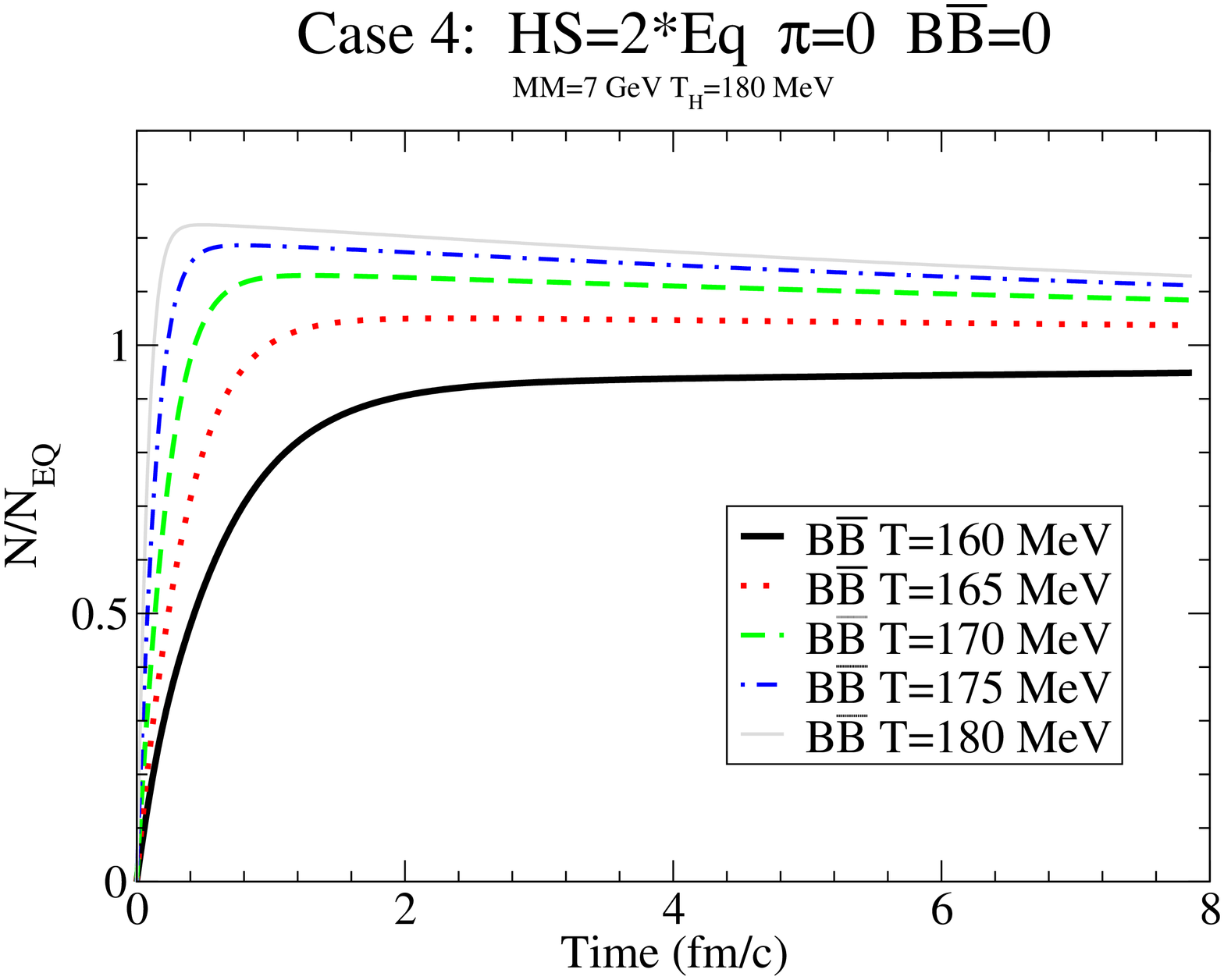,width=0.3\linewidth,angle=270,clip=} \\
\end{tabular}
\caption{Solution of the rate equations when the resonances, pions,
and baryon anti-baryon pairs are allowed to change over time.}
\label{fig:HS2pi0bab0}
\end{figure}

When everything is allowed to develop out of equilibrium,
quasi-equilibrium is reached on a time scale of the previous
estimated equilibration times.  The results are given in Fig.\
(\ref{fig:HS2pi0bab0})
Although we only see a small deviation in the pions from equilibrium in Fig.\ (\ref{fig:HS2pi0bab0}) that minor deviation drastically affects the resonances because even the lightest resonance, whose mass is $M=2$ GeV, decays on average into $\langle n\rangle\approx 5$ pions. In Fig.\ (\ref{fig:HS2pi0bab0}) we clearly see that quasi-equilibrium is reached quickly i.e. $\tau_{quasi-eq}<1\;\frac{\mathrm{fm}}{c}$.  After quasi-equilibrium is reached the remaining constituents (especially the resonances) slowly reach equilibrium.  What we can get from Fig.\ (\ref{fig:HS2pi0bab0}) is that the pions and the baryon anti-baryon pairs quickly equilibrate, especially for higher temperatures.  In the graph of the baryon anti-baron pairs we see that at $T=180$ MeV the baryon anti-baryon pairs quickly populate near chemical equilibrium and while they are not in complete equilibrium are quite close to it. Hence, for temperatures between $T=180\;\mathrm{MeV\;and}\;160$ MeV, the baryon anti-baryon pairs are populated with equilibration times faster or equal to $2\;\frac{\mathrm{fm}}{c}$.  This constitutes a very promising finding.

\section{Conclusions}

Our preliminary results and time scale estimates indicate that baryon anti-baryon pairs can be ``born" out of equilibrium  after hadronization and then equilibrated by the subsequent population and decay of Hagedorn states.  Cases 1-3 for the decay $HS\leftrightarrow n\pi+B\bar{B}$ clearly show quick equilibration times for baryon anti-baryon pairs  between temperatures of $180$ MeV and $160$ MeV i.e. slightly below the phase transition.  When all the particles started out of equilibration the baryon anti-baryon pairs quickly neared equilibrium although a minor deviation was still observed from chemical equilibration.  Afterwards, due to affects from the need to kill of the number of effective pions in the system, longer time scales were observed when the pions, Hagedorn states and baryon anti-baryon pairs were out of equilibrium.  However, the pions and baryon anti-baryon pairs were quickly populated near $T_{c}$ and remained close to their equilibrium values even when it took longer for the Hagedorn states to reach chemical equilibrium.  Since the Hagedorn states only contribute near $T_{c}$ these appear to be acceptable results.

In an upcoming publication we will delve more thoroughly into the effects of the quasi-equilibrated state seen in both Fig. (\ref{fig:2re0pi}), (\ref{fig:HS2pi0bab0}) and we also will discuss its effects on the fireball as it cools over time due to a Bjorken expansion.  Thus, we expect to see the baryon anti-baryon pairs quickly reach equilibrium and then they will not be produced further when the system is cooled.  Moreover, we will vary our choices of initial conditions and discuss their implications.  Finally, a non-zero strangeness will be included specifically in the baryon anti-baryon pairs so that we can study the equilibration times of anti-hyperons and multi-strange baryons.  The continuation of our work is promising in order to explain results for strangeness in baryons and anti-baryons found at RHIC.  
\section{ACKNOWLEDGEMENTS}
This work was supported by FIGSS. J.N-H. would also like to
thank Jorge Noronha for thoughtful discussions on this work.  Furthermore, it is a pleasure to thank the organizers of the XLV International Winter Meeting on Nuclear Physics, Bormio 2007 for allowing us to present our results.

\end{document}